\documentclass[twocolumn,showpacs,preprintnumbers,amsmath,amssymb]{revtex4}

\usepackage{graphicx}
\usepackage{dcolumn}
\usepackage{bm}

\begin{document}

\preprint{APS/123-QED}

\title{Chaotic Synchronization Via Minimum Information Transmission}

\author{A. S. Dmitriev}
\author{G. A. Kassian}%
\author{A. D. Khilinsky}
\affiliation{%
Institute of Radioengineering and Electronics Russian Academy of
Sciences\\
Mokhovaya 11-7, Moscow, 101999, Russia}%

\author{M. Hasler}
\affiliation{ Ecole Polytechnique Federale de Lausanne, Lausanne,
Switzerland}%

\date{\today}

\begin{abstract}
Chaotic synchronization is generally extremely sensitive to the
presence of noise and other inference in the channel. Is this
sensitivity a fundamental property  of chaotic synchronization or
is it related to the choice of synchronization method and  can it
be suppressed by a modification of the method? If the answer is
positive, then what are the relationships between the properties
of a dynamical system and the level of noise at which the
suppression of this sensitivity is still possible? What are
particular methods to achieve synchronization  stable to the
presence of noise? In this paper we present the analysis of this
issue from the standpoint of information theory. The fundamental
reason for this sensitivity is the fact that the chaotic signal
contains information which requires a certain minimal threshold
signal-to-noise ratio for transmission. Only in this case
high-quality synchronization is achievable and only if the
required information is transmitted (coded) optimally. Otherwise
the threshold level can be much higher.
\end{abstract}

\pacs{05.45.-a,07.50.Qx,95.75Wx}
\maketitle

\section{\label{sec:level1}Introduction}
The possibility to synchronize the dynamics of  two identical
chaotic systems \cite{1,2,3,4}  is determined by  two factors (to
be precise the case of unidirectional impact of one system on the
other is assumed, i.e., the situation "drive system - response
system" is considered). First, the way the drive system affects
the response system. Significant progress has been achieved here
that enables synchronization not only of chaotic systems but also
of systems with hyperchaos \cite{5,6}. Second, it is determined by
the sensitivity of the chaotic synchronization to distortions of
the signal in the coupling channel, in particular, to the effects
of filtering and to the presence of noise \cite{30}. The attempts
to damp this sensitivity are being made both in the direction of
development of more robust schemes \cite{29,23} and in the
direction of correction (reconstruction) of a signal, distorted in
the coupling channel, into its original form at the input of the
responce system. For instance, in the case of a noisy signal, the
reconstruction is considered to be the cleaning of this signal
from noise \cite{7,8,9,10,27,31}. Hence, the problem is to
reconstruct with a high precision the output signal of the drive
system at the input of the response system. However, for this
purpose, transmission of the signal itself through the coupling
channel is not necessary. Instead, one can transmit the minimal
information about the signal necessary to reconstruct (recover)
the initial signal at the input of the response system, i.e. to
replace the coupling channel by the communication channel. From
the viewpoint of the information theory  the second approach to
reconstruction of the signal at the input of the response system
can be preferable as it (i) requires less information to be
transmitted and (ii) this information can be presented (coded) so
that it is more robust to the disturbances in the communication
channel.

This idea was discussed in the literature \cite{12,13} and was
tested in the case when the chaotic systems were one-dimensional
maps \cite{13}. Here, on the example of two-dimensional maps we
show that this approach can be used for synchronization  of a
higher-dimensional chaotic maps (both hyperbolic and
non-hyperbolic).

As an example of two-dimensional maps the hyperbolic Lozi map
\cite{14} is used in this paper. It is described  by the equations
\begin{eqnarray}\label{eq1}
x_{n+1}&=&(\alpha-1)-\alpha|x_n|+y_n,\nonumber\\
y_{n+1}&=&\beta x_n,
\end{eqnarray}
where $\alpha$ and $\beta$ are  parameters.

Non-hyperbolic maps are represented here by the Henon map
\cite{15}
\begin{eqnarray}\label{eq1a}
x_{n+1}&=&1-\alpha x_n^2+y_n,\nonumber\\
y_{n+1}&=&\beta x_n,
\end{eqnarray}

\section{Restrictions from Information Theory on Synchronization}

Drive chaotic system is a specific source of information.  With
respect to information transmission, the driven system can be
treated as a receiver of information, that, if synchronization
takes place, receives all the information defining the state of
the dynamical system, that is sent through the communication
channel and continuously  adjusts its dynamics in accordance with
the information received.

Characteristic of a chaotic source are a finite rate of
information production \cite{16,17} and continuous signal values.
A question arises, what the necessary carrying capacity of a
communication channel between the two systems is and how it
relates to the possibility of high-quality synchronization.

According to Shannon's theorem \cite{18}, in order to transmit
without errors information produced by the source at rate $I$
through a channel of capacity $C$ one must have
\begin{equation}
\label{eq2} C>I.
\end{equation}
Hence, for synchronization of the drive and response systems in
the presence as well as in the absence of noise it is sufficient
to have a communication channel with a carrying capacity larger
than $I$.  This fact has been noticed in \cite{12}. It suggests
that high-quality synchronization can be achieved, if external
perturbations do not exceed some critical value.

The theorem also lays  the basis for the quantitative analysis of
the capacity of a noisy channel. For a channel  with additive
white gaussian noise, and bandwidth $W$ the  capacity per unit of
time is
\begin{equation}
\label{eq3}
 C=W\log_2\frac{P+N}{N}.
\end{equation}
where $P$ is the signal power and $N$ is the noise power.

Taking the sampling period at $T=\frac{1}{2W}$ so that the whole
bandwidth is used, capacity per sample becomes

\begin{equation}
 C=\frac12 \log_2\frac{P+N}{N},\label{eq3a}
\end{equation}

Eq.~(\ref{eq3}) implies that when appropriate coding is used,
information can be transmitted with the rate up to $W\cdot
\log_2\frac{P+N}N$ bit/sec while error probability can be
arbitrarily low.

Equation (\ref{eq3}) also suggests that if the information
production rate $I$ and the channel capacity are known, the
maximum noise level can be calculated such that high-quality
synchronization  is still feasible.

As an example, if $I$ = 1 bit/sample, then, in theory, arbitrarily
precise synchronization can be achieved when the ratio of the mean
squared power of the chaotic signal $P=<x^2_n>$ to mean squared
power of gaussian  noise $N=<w^2_n>$ is
$$P/N>3.$$
The "expense" of the possibility of high precision synchronization
in the presence of noise is a special organization of the
information transmitted through the channel and a certain time
delay of the processes in the response system with respect to the
drive system. Note also, that for the practical accomplishment of
high-quality synchronization  in the presence of noise a certain
increase of the channel capacity is necessary compared  to the
value given by relation (\ref{eq2}). Otherwise, the delay value
will tend to infinity.

\section{Hyperbolic Case. Lozi Map.}

So, in order to be able to synchronize the response system with
the drive system we should transmit information at a certain rate.
Let us consider the information production of the Lozi map and the
amount of information that must be transmitted for
synchronization.

First, let us recall that for 1-D maps
\begin{equation}
\label{eq7} x_{n+1}=f(x_n)
\end{equation}
the average information production $I$, measured in bits per
sample, can be expressed as the integral of (\ref{eq7}) weighted
with the invariant probability density $P(x)$
\begin{equation}\label{eq8}
    \int_{0}^{1}P(x)\log_2\left|\frac{df}{dx}\right|dx.
\end{equation}

Relation (\ref{eq8}) coincides with the definition of the Lyapunov
exponent $\lambda$ of the one-dimensional map with the only
difference that in the relation for the Lyapunov exponent the
natural logarithm is used instead of the base-2 logarithm. Hence,
the Lyapunov exponent for 1-D map can be treated as the
information production rate expressed in base-$e$  units and it is
equivalent to the Kolmogorov entropy $h$. In order to transform it
to bit/sample, we multiply $\lambda$ by $\log_2e$. Let us remark
that the Kolmogorov entropy $h$ and the information production
rate $R$ are generally the same not only in the case of 1-D maps,
but also in more general cases. However, it is common to express
$R$ in bits/sample and $h$ in nats/sample.

The situation with information production in 2-D maps, e.g. in the
Lozi map, is more complicated. It is well known \cite{17} that for
differentiable maps the following inequality between Kolmogorov
entropy $h$ and positive Lyapunov exponents holds:
\begin{subequations}\label{eq10}
\begin{equation}
h\leq\lambda_1+\lambda_2,
\end{equation}
if both  Lyapunov exponents are positive, or
\begin{equation}
h\leq\lambda_1,
\end{equation}
\end{subequations}
if only $\lambda_1$ is positive. Under some additional conditions
inequalities (\ref{eq10}) can turn into equalities. There are no
similar results for the non-differentiable Lozi map. However, one
can try to use another approach to estimate the information
production by the Lozi map.

Let $N_n$ be the number of unstable periodic orbits of length $n$
in the chaotic attractor, and thus $L_n=N_n\cdot n$ the number of
cyclical points. Then the following limit is supposed to estimate
topological entropy
\begin{equation}\label{eq11}
K_0=\lim_{n\to\infty}\frac1n\ln L_n,
\end{equation}
Estimates for information production $I_n$, calculated with the
aid of (\ref{eq11}) are presented in Table~ {\ref{table1}}  for
different $n$.
\begin{table}
\caption{\label{table1}Periodic orbits of Lozi map attractor.
$(\alpha=1.7, \beta=0.5)$; $n$ - the length of a cycle; $N_n$ -
the number of cycles with the length n; $L_n$ - the number of
periodic points; $I_n$ - information production estimate.}
\begin{tabular}{|c|c|c|c|c|c|c|}
  \hline
  $N$ & 1 & 4 & 8 & 12 & 16 & 17 \\
  \hline
  $N_n$ & 1 & 1 & 8 & 34 & 162 & 224 \\
  \hline
  $L_n$ & 1 & 4 & 64 & 408 & 2592 & 3808 \\
  \hline
  $I_n$ & 0 & 0.5 & 0.75 & 0.722 & 0.720 & 0.700 \\
  \hline
\end{tabular}
\end{table}
Both methods of estimating information production (including the
method based on the positive Lyapunov exponents ) give similar
results.

\begin{figure}
  \includegraphics[width=9cm]{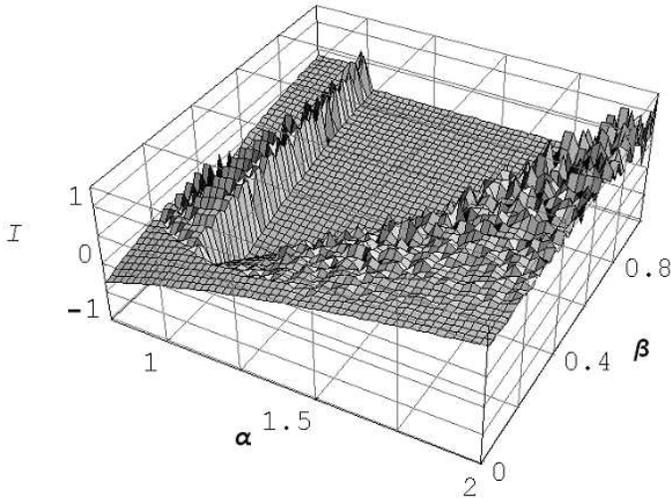}
  \caption{Information production of Lozi map as a function of parameters $\alpha$ and $\beta$.}\label{fig3}
\end{figure}

Fig.~\ref{fig3} shows the average information production for
different values $\alpha$ and $\beta$. For $\alpha=1.7$;
$\beta=0.5$ the average information production rate is
$I\approx0.7$ bits/sample.

\begin{figure}
  \includegraphics[width=9cm]{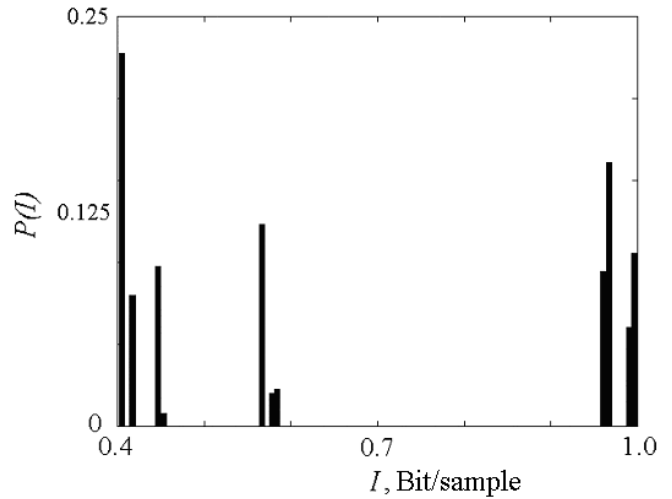}
  \caption{The probability distribution function of the rate of information production of the Lozi map.}\label{fig4}
\end{figure}

However, information production differs from iteration to
iteration. Fig.~\ref{fig4} shows the information production rate
probability density function. As  can be seen in the figure the
rate never exceeds 1 bit/sample.

Consequently, to ensure synchronization a channel with capacity
$C>I=0.7$ bits/sample can be used. However, it is desirable that
the channel capacity should be at least $C=I_{max}=1$ bit/sample
in order to take the maximum information production of each sample
into account. Below a symbolic sequence for Lozi map is studied as
an example of such a binary sequence.

\textbf{Synchronization scheme}

\begin{figure}
  \includegraphics[width=9cm]{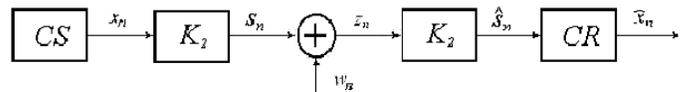}
  \caption{Synchronization by means of symbolic sequence. $CS$-chaotic source. $CR$-chaotic receiver, $K_2$-quantizer.}
  \label{fig5}
\end{figure}

 The proposed scheme is shown in
Fig.~\ref{fig5}. The signal $x_n$ produced by iteration of the
Lozi map is transformed by a 2-level quantizer (threshold device)
into a symbolic sequence $X_n=\text{sign}(x_n)$, i.e. into -1 if
$x_n<0$ and into +1 if $x_n\geq0$. Then this binary signal is sent
through the channel, where it is corrupted by noise:
$z_n=X_n+w_n$. The quantizer transforms it back to the binary
signal $\hat{X}_n=\text{sign}(z_n)$.

The $\pm1$ values of binary sequence correspond to symbols $L$ and
$R$ of the symbolic sequence. This symbolic sequence is used in
the reconstruction block to restore the chaotic sequence. Thus, in
the absence of noise, chaotic sequence reconstructed with a high
precision is fed to the input of the response system, or an
estimate of this chaotic sequence, if there is noise.

First, a problem with the noisefree case will be analyzed.

\textbf{Reconstruction of a Trajectory from Symbolic Sequence}

According to the synchronization scheme, at the input of the
response system the incoming symbolic sequence must be used to
reconstruct the chaotic sequence generated by the chaotic source.

In the case of one-dimensional maps reconstruction of the chaotic
sequence based on the symbolic sequence can be accomplished by
iterating the map backward \cite{12,13}. Since the map is
stretching, being iterated backward it becomes contracting. Hence,
repeatedly applying the inverse map to the initial conditions
given by the elements of the symbolic sequence can give a good
estimate of the chaotic sequence.

Consider a procedure of reconstruction of a chaotic sequence
 given the symbolic sequence for the case of Lozi map.

The Jacobian of (\ref{eq1}) is constant and equals $\beta$. This
means that the area of a small cell will be contracted by each
iteration of the Lozi map by a factor $\beta$. But the distance
between two close points increases in a certain direction, the
stretching direction, at a rate determined by the positive
Lyapunov exponent $\lambda_1$. Let two points in the plane be
separated along the stretching direction by the small distance
$\delta_0$ at time 0. Then the distance between the corresponding
trajectories at moment $m$ is approximately
\begin{equation}\label{eq16}
\delta x_m\propto\delta_0\cdot\exp(\lambda_1\cdot m),
\end{equation}
The situation is very similar to the case of the 1-D map, where
one could use this property (\ref{eq16}) of the map to improve our
knowledge of the value of the trajectory at present on the basis
of information about its values in the future. Namely, to improve
the precision of $x_n$ the map should be iterated in reverse time
beginning from $x_{n+k}$.

But what happens if we iterate the Lozi map in reverse time? In
this case the area of a small cell will not decrease but increase.
The direction corresponding to the positive Lyapunov exponent will
be contracting but the direction corresponding to the negative
Lyapunov exponent will be stretching. The trajectories will
diverge again. The rate of this divergence will be determined by
the absolute value of the second Lyapunov exponent.

The fact that two trajectories will diverge  when the map is
iterated either forward or backward can be written as
\begin{equation}\label{eq17}
\delta x_m\propto\delta_0\cdot\exp(\lambda_1\cdot
m)+\varepsilon_0\cdot\exp(\lambda_2\cdot m),
\end{equation}
where $\varepsilon_0$ is the initial distance between the two
points in the direction of contraction. As follows from
(\ref{eq17}), in order to improve the knowledge about the position
of the trajectory at the present time, it is necessary to use
information both about the "future" and the "past" of the
trajectory.

We rewrite equations (\ref{eq1}) for the Lozi map in such a way as
to make it contracting both from the future and from the past:
\begin{equation}\label{eq18}
|x_n|=\frac{\alpha-1}{\alpha}-\frac1\alpha
x_{n+1}+\frac\beta\alpha x_{n-1}.
\end{equation}

Map (\ref{eq18}) is a contracting map because coefficients
$1/\alpha$ in front of $x_{n+1}$ and $\beta/\alpha$ in front of
$x_{n-1}$($\alpha=1.7, \beta=0.5$) as well as their sum are less
than $1$. Contracting features of map (\ref{eq18}) may be used to
improve an estimate of the variable value at the $n$-th moment
 using information about the variable values at the ($n-1$)-th and
($n+1$)-th moments. Note that map (\ref{eq18}) is not
single-valued but a double valued map.

\textbf{Reconstruction algorithm.} So, let us have a symbolic
sequence $X_1,\ldots X_n,\ldots X_{n+k},\ldots X_N$, where $X_k\in
\{-1,1\}$: $N$ is the number of elements in the sequence. We
evaluate $|x_n|$ at the  first step for all elements besides the
first and the last, i.e. for $n=2,\ldots,N-1$  in the following
way
\begin{equation}
x_n^1=X_n|x_n^1|=X_n\left[\frac{\alpha-1}{\alpha}-\frac1\alpha
X_{n+1}+\frac\beta\alpha X_{n-1}\right].
\end{equation}

Elements of the symbolic sequence are used at this stage as
initial approximation for the estimate. At the $i$-th step of the
procedure the estimates obtained at the previous ($i$-1)-th step
are used.
\begin{equation}
x_n^i=X_n|x_n^i|=X_n\left[\frac{\alpha-1}{\alpha}-\frac1\alpha
x_{n+1}^{i-1}+\frac\beta\alpha x_{n-1}^{i-1}\right].
\end{equation}
where $n=2,\ldots,N-1$. We finish the process when the difference
between estimates of ${x^i}$ on  $i$-th and $i+1$-th iterations is
less then some small $\varepsilon$.

\begin{figure}
\includegraphics[width=9cm]{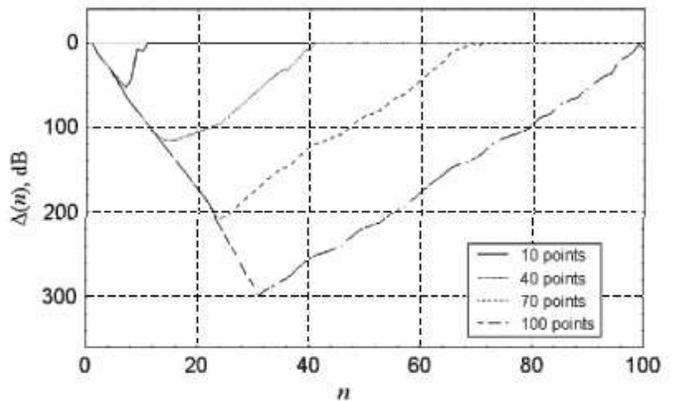}
\caption{Reconstruction of chaotic sequence by means of symbolic
sequence. $n$ --- reconstructed point index, $\Delta(n)$
--- reconstruction error, $N=10,40,70,100$ --- the number of
elements of the symbolic sequence that are used for
reconstruction.}\label{fig8}
\end{figure}

The results of  chaotic sequence reconstruction  using the above
algorithm are presented in Fig.~\ref{fig8}. The four curves in
Fig.~\ref{fig8} correspond to the relative precision (in dB)
obtained by taking  $N=10,40,70\text{ and }100$ points of the
symbolic sequence. Each of these curves looks as two sides of a
triangle.

Consider the longest curve, that shows the results of the
reconstruction of a fragment  100 points long. It can be seen that
the first and the last points are not reconstructed at all, they
are just the values of the symbolic sequence. The precision at
these points corresponds to the precision of the initial
approximation.  The best result of reconstruction with precision
$\sim10^{-15}$ corresponds to the region of 30-th point. The ratio
of the number of points to the left and to the right from this
point is 3:7 or 1:2.3 that is approximately the ratio of the
coefficients $1/\alpha$ and $\beta/\alpha$ in (\ref{eq18}):
$\beta/\alpha:1/\alpha=1:2$

Hence, to reconstruct the sequence of chaotic signals $x_1,\ldots
x_n$ with a given precision  , one first has to calculate the
length of the fragment of the symbolic sequence to be used for
reconstruction of one point  of the chaotic sequence. For
instance, the length of the fragment equals 100 when the
reconstruction precision is $10^{-15}$. Then, the left-edge
fragment of the symbolic sequence $X_1,\ldots X_{100}$ is used to
reconstruct the 30th point of the chaotic sequence $x_{30}$. Then
the fragment $X_2,\ldots X_{101}$ is used to reconstruct the point
$x_{31}$ of the chaotic sequence and so on. As for reconstruction
of any given point of the chaotic sequence one needs 29 preceding
and 70 following points of the corresponding symbolic sequence.
Reconstruction can be accomplished with this precision only for
those points that are non less then 29 positions distant from the
left edge and 70 positions distant from the right one.

\section{Non-hyperbolic Henon Map}

In the case of non-hyperbolic maps, e.g. the Henon map, the
approach based on the transmission of the symbolic sequence with
subsequent reconstruction of the chaotic sequence fails to be
directly applicable.  However, an attempt to apply this approach
allows to understand why the results are insufficient and helps to
develop a suitable method.

Assume first, as  in the case of the Lozi map, that a sequence
$X_n=\text{sign}(x_n)$ is transmitted through the channel. Note
that the sequence $x_1,...x_N,...$ is not the symbolic sequence
for this case.

Henon map can be rewritten as a one-dimensional map of the second
order
\begin{equation}\label{eqhenon5a}
x_{n+1}=1-\alpha x_n^2 + \beta x_{n-1}.
\end{equation}
When iterated forward, this map is stretching at almost all points
of the trajectory, that is, the error in the value of $x_{n+1}$
usually is greater than that of $x_n$. On average, the distance
between two close points increases exponentially with time.

Rewrite (\ref{eqhenon5a}) in the form
\begin{equation}\label{eqhenon5}
x_n=\frac1{\sqrt{\alpha}}\sqrt{1-x_{n+1}+\beta x_{n-1}}.
\end{equation}

\begin{figure}
\includegraphics[width=9cm]{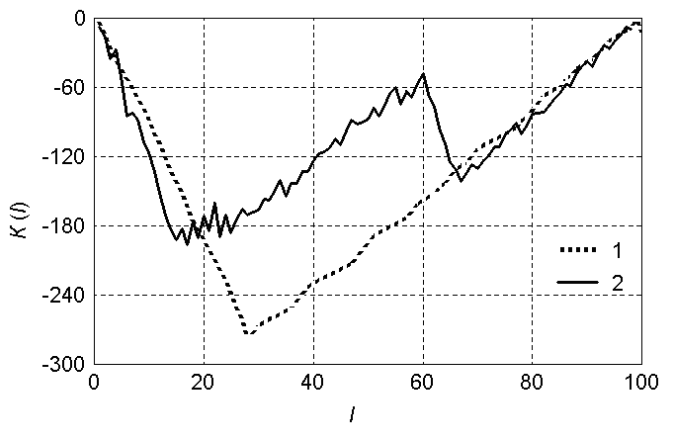}
\caption{Reconstruction of chaotic sequence by means of symbolic
sequence. $l$ --- the number of reconstructed point, $k(l)$
--- reconstruction error, the symbolic sequence is reconstructed from 100 sample
points. 1 --- application of the algorithm to Lozi map, 2
--- application of the algorithm to the Henon map}\label{fighenon1}
\end{figure}

Map (\ref{eqhenon5}) is contracting everywhere but in the region
of small $x_n$: $|x_n|<\varepsilon$, where $\varepsilon \sim 0.1$.
Outside this region the error of the estimated value of $x_n$
according to (\ref{eqhenon5}) is smaller than the error of the
estimates of $x_{n-1}$ and $x_{n+1}$ that are used to get an
estimate of $x_n$. For small $|x_n|$ this is not true, hence, the
algorithm analogous to the one used for the Lozi map can not be
used to reconstruct the chaotic sequence with high precision. A
typical result of the application of this algorithm to the Henon
map is shown in Fig.~\ref{fighenon1} (line 2). One can see that
there are intervals with  sharp reduction of reconstruction
precision and the intervals where the precision of reconstruction
is relatively high. Deterioration of the precision always occurs
because of the lack of contracting properties of map
(\ref{eqhenon5}) for small $|x_n|$.

Assume that  at the reconstruction block the estimates of a
chaotic sequence at the moments  $n_0$ and $n_0+1$ are known with
the desired precision and  there is also a copy of the Henon map.
Then this information is sufficient to get the values of the
elements of the binary sequence $X_n$ for several points with the
indices $n>n_0$. For this one just have to iterate the Henon map
with initial conditions $(x_{n_0},x_{n_0-1})$ and construct a
"forecast". This "forecast" allows to determine the moments of
time when the trajectory goes through a region of small absolute
values.

Since initial conditions usually are known only with a finite
precision, the correct values for the elements of the symbolic
sequence can be obtained only for a relatively short part of the
trajectory when the error of the "forecast" is much less then the
size of the attractor.

Above, in the description of the algorithm for  Lozi map  it has
been noted that for a reconstruction of a chaotic sequence
$x_1,\ldots x_N$ with a high quality precesion  it is sufficient
to know the symbolic sequence $X_n$ for $n\in[2,N-1]$ and
$x_1,x_N$ with the desired precision. The same conditions hold for
the case of Henon map given that in the range $n\in[2,N-1]$ there
are no $x_n$ with absolute values close to zero.

\begin{figure}
\includegraphics[width=9cm]{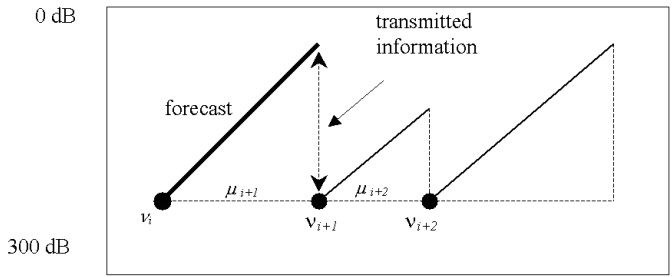}
\caption{Forecast of chaotic trajectory and transmission of
necessary information for forecast amendment.}\label{fighenon2}
\end{figure}

Consequently, we arrive at the following algorithm of the
information transmission from the drive to the response system.
Let $\nu_i$ be the indices of the points where the trajectory
comes close to zero, $\mu_i=[\nu_i,\nu_{i+1}]$ - the segment of a
trajectory limited by two subsequent points close to zero (see
Fig.~\ref{fighenon2}). Assume that at the input of the driven
system there is an estimate of the chaotic sequence for all points
of $\mu_i$. The two last points from $\mu_i$ are taken as the
initial conditions for the forecast of points from $\mu_{i+1}$.
Note also that the length of $\mu_{i+1}$ is initially unknown at
the response system side. Since the initial conditions for a
forecasting trajectory are known with some error, a disturbed and
undisturbed trajectories diverge exponentially and the precision
of the forecast  also grows exponentially with the distance from
$\nu_{i+1}$. Nevertheless, if the length of $\mu_{i+1}$ is not
very large, one can: (1) predict the index of  point $\nu_{i+2}$
where the chaotic trajectory is close to zero and (2) construct
the binary sequence $X_n$ for all points from $\mu_{i+1}$.

At this stage the only single datum  needed for chaotic sequence
reconstruction is the value  at the point $\nu_{i+2}$. This value
is transmitted through the channel with the necessary precision.
After that $\mu_{i+1}$ is reconstructed with the aid of the
algorithm described for the Lozi map. Hence, to reconstruct the
chaotic sequence at all points of $\mu_{i+1}$ one only has to
transmit the value of a single point (with the index $\nu_{i+2}$).

A subtle modification of this algorithm can substantially reduce
the amount of information transmitted. The case is that the
forecast for the points from $\mu_{i+1}$ allows to get some
estimate for the point $x_{\nu_{i+2}}$, and so, there is no need
to transmit the whole value of this point. One just needs to
transmit the bits that can be distorted due to exponential
divergence of the trajectories. Since the rate of divergence is
solely determined by the first Lyapunov exponent, the amount of
information to be transmitted through the communication channel is
also determined only by this property of the chaotic map and is
independent of the desired reconstruction precision of the chaotic
sequence.

A numerical simulation of the proposed algorithm  ensures  high
precision of synchronization with a volume of transmitted
information $\sim1$ bits/sample, that is close to the rate of
information production by Henon map ($\sim~0.6$ bits/sample). Note
that in \cite{24} the method of synchronization  with dynamical
coupling was analyzed that allows to reduce the amount of
necessary information  to be transmitted for synchronization by a
factor of 5-6 as compared to conventional procedure. The approach
considered above provides a reduction of information flow by a
factor of 20-30 in comparison with conventional procedure.

\section{Synchronization  in the Presence of Noise}

It has been shown  that in order to achieve synchronization of
hyperbolic as well as non-hyperbolic maps it is sufficient to
transmit information contained in the chaotic sequence and then
utilize this information to reconstruct the chaotic sequence. The
precision of reconstruction depends on the length of the sequence
and on the number of elements of the symbolic sequence that are
directly used  to reconstruct the chaotic sample. The
reconstructed sequence of samples ensures high-quality
synchronization of  drive and response systems.

If binary information is transmitted via a noisy channel errors
may appear in the binary sequence. These errors may lead to
incorrect reconstruction of the chaotic signal and thus reduce the
synchronization precision.

Consider the results of numerical simulation of the precision of
synchronization  of chaotic signal reconstruction after
transmission through a noisy channel. Note that errors can be made
arbitrarily small by increasing the channel capacity (e.g., by
means of using error correcting codes). Hence, synchronization
through channels with a capacity close to the minimum is of
greater concern.

\begin{figure}
\includegraphics[width=9cm]{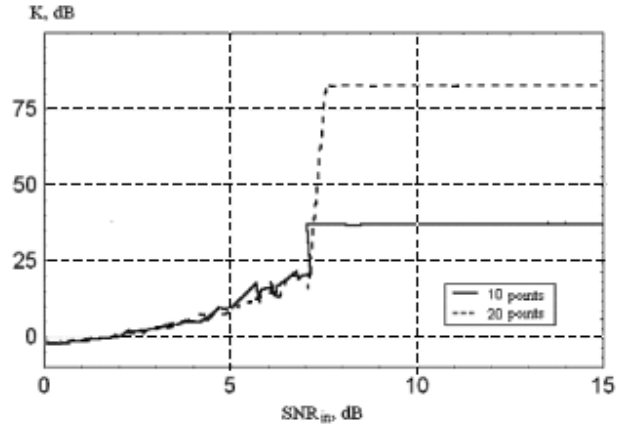}
\caption{Synchronization quality for the case of AWGN channel.
$K=(SNR_{in}-SNR_{out})$ --- the criterion of the synchronization
quality.}\label{fig10}
\end{figure}

Fig.~\ref{fig10} shows graphs that illustrate the quality of
synchronization when information is transmitted through a channel
with gaussian noise.

It can be seen that for signal to noise ratios up to about 7.5 dB,
the synchronization error is actually given by the noise and it
cannot be improved by increasing the number of points used for the
reconstruction of a sample. Above this threshold, the
synchronization error is decreased by about 4.2 dB/points used in
the reconstruction, which is in good agreement with the rate of
information production $I$ of 0.7 bit/sample.

Similar results were obtained for the synchronization of Henon
maps.

\section{Conclusion}
In this paper  a synchronization scheme for two-dimensional maps
is considered that is based on  transmission of  information about
the state of drive system by unidirectional coupling. It is shown
that the necessary amount of information that must be delivered to
the response system  is determined by the rate of information
production by the dynamical system.

When this threshold is increased, however, arbitrarily precise
synchronization can be achieved. The only price to pay for that is
the complexity of the reconstruction algorithm and the delay of
the reconstructed signal. This is contrary to the common belief
that the synchronization error necessarily monotonically increases
with the noise.

The schemes of synchronization of  two-dimensional maps are
presented, that are based on  reconstruction of the chaotic
sequence. Their stability with respect to noise almost coincides
with the theoretical estimate of the ultimate limit.

\begin{acknowledgments}
This work has been supported by the Swiss National Science
Foundation, SCOPES project Nr. 75SUPJ062310.
\end{acknowledgments}

\newpage 
\bibliography{pre}

\end{document}